\documentclass{article}

\title{An Analysis of Stapp's ``A Bell-type theorem without hidden
variables''}

\author{Abner Shimony
\\
Departments of Philosophy and Physics
\\
Boston University, Boston, MA}

\date{}

\begin{document}

\maketitle

\begin{abstract}
H.P. Stapp has proposed a number of demonstrations of a Bell-type
theorem which dispensed with an assumption of hidden variables, but
relied only upon locality together with an assumption that
experimenters can choose freely which of several incompatible
observables to measure. In recent papers his strategy has centered
upon counterfactual conditionals. Stapp's paper in American Journal
of Physics, 2004, replies to objections raised against earlier
expositions of this strategy and proposes a simplified demonstration.
The new demonstration is criticized, several subtleties in the logic
of counterfactuals are pointed out, and the proofs of J.S. Bell and
his followers are advocated.
\end{abstract}

\section{Introduction}

Henry Stapp's article ``A Bell-type theorem without hidden
variables''~\cite{1} is a reformulation of a project which has
evolved over many years~\cite{2,3,4,5,6,7,8} and has been the subject
of numerous critical assessments, e.g., Clauser and Shimony~\cite{9},
Clifton, Butterfield, and Redhead~\cite{10}, Clifton and
Dickson~\cite{11}, Mermin~\cite{12}, Unruh~\cite{13}, Shimony and
Stein~\cite{14,15}. The last four of these criticisms were directed
specifically against Stapp~\cite{6}. Concerning these four criticisms
Stapp~\cite{1} states on column 1 of p.\ 32, ``I have answered these
objections. However, the very existence of these challenges shows
that the approach used \ldots\ has serious problems, which originate
in the fact that it is based on classical modal logic.'' The purpose
of the present paper is to assess carefully Stapp~\cite{1}, restating
previous criticisms when his new language leaves the scientific and
philosophical theses of  Stapp~\cite{6} essentially unchanged,  but
also acknowledging and assessing innovations. Stapp's seriousness and
tenacity merit an attentive examination of his latest exposition.

Stapp~\cite{1,6}  expresses dissatisfaction with the alleged
demonstrations of  J.S. Bell~\cite{16,17} and his followers~\cite{18}
that the predictions of Quantum Mechanics (QM) are inconsistent with
the locality of the Special Theory of Relativity (STR). All of these
demonstrations examine the possibility of recovering the predictions
of QM by means of a local hidden-variables model, which assigns
definite values to all the observable quantities recognized by QM
even when QM prohibits in principle their simultaneous measurability.
In thus characterizing the entire class of local hidden variables
theories Stapp argues that the difference between so-called
``deterministic'' and so-called ``stochastic'' theories is
superficial, because the latter are equivalent to the former except
for errors which tend to zero as the number of experiments goes to
infinity; see especially Stapp~\cite{3}. Consequently, Stapp
maintains, the demonstrations by Bell and his followers prove only
the inconsistency of QM with the {\it conjunction\/} of locality,
free choice of experiments, and the assumption of hidden variables,
and since the assumption of hidden variables violates the
philosophical viewpoint of QM, the conclusion reached by the standard
derivations is neither surprising nor profound. Stapp aims at a
stronger theorem throughout his entire series of investigations of
the foundations of quantum mechanics: that {\it QM is inconsistent
with the conjunction of locality and the assumption  that
experimenters freely choose among incompatible experiments
performable in a given space-time region}, the latter being an
assumption which he regards as implicit in the entire enterprise of
experimental science. I wish to assert, at the commencement of my
critical assessment,  my agreement with Stapp that a successful
demonstration of the inconsistency of QM with locality and free
choice, without the explicit or implicit assumption of hidden
variables, would indeed be a profound scientific and philosophical
achievement.

\section{Stapp's Explicit and Tacit Assumptions}

The physical system S which Stapp envisages for analysis consists of
two parts, located in two space-time regions {\bf R} and {\bf L},
which have space-like separation from each other. Einstein-Minkowski
geometry, which is the geometry of the space-time of the Special
Theory of Relativity (STR) implies that there exists a Lorentz frame
F in which every point in {\bf L} is earlier than every point in {\bf
R}, and Stapp prefers to speak of {\bf R} and {\bf L} in the frame F
whereas I prefer to express the relation between {\bf R} and {\bf L}
in a frame-invariant manner, even though the same geometrical facts
can be equivalently expressed in both ways. The part of S located in
{\bf L} will be called $\lambda$, and the part located in {\bf R}
will be called $\rho$. $\lambda$ is subjected to one of two
measurements, labeled L1 and L2, freely chosen by an experimenter in
{\bf L}, while $\rho$ is subjected to one of two experiments, labeled
R1 and R2, freely chosen in {\bf R}. Each measurement is bivalent,
with possible results labeled $+$ and $-$. The expression ``L1$+$''
ambiguously designates the $+$ result of the measurement L1 or the
proposition that the measurement L1 is performed with outcome $+$;
and ``L1$-$'', ``L2$+$'', and ``L2$-$'' have analogous meanings.
Context determines which of the possible meanings is intended.
Likewise each of R1 and R2 has bivalent outcomes $+$ and $-$.
Following L. Hardy~\cite{19} and the slight modification by Eberhard
and Rosselet~\cite{20} (the latter referred to here henceforth as
``HER'', but Stapp calls it ``Hardy''), Stapp ascribes to the
composite system a quantum state which predicts the following
correlations:
\begin{enumerate}
\item
Upon the condition that L2 is performed, if R2$+$ then L2$+$.
\item
If R1 is performed and L2$+$, then R1$-$.
\item
If R2 is performed and L1$-$, then R2$+$.
\item
If  R1 is performed and L1$-$, then  R1 has a non-zero probability of
having outcome $+$ and a non-zero probability of having outcome $-$.
(Actually HER choose these probabilities to be 50\%, but I prefer not
to work with quantitative probabilities for reasons to be stated
later.
\end{enumerate}

Stapp considers probability to be a meaningful concept applicable to
a single system, like the system S, even though empirical evidence
about probabilities can be gathered only when one has an ensemble of
similar systems, all prepared in the same quantum state. In the
analysis of counterfactual conditionals at the heart of Stapp's
theorem a central concept is that of ``possible worlds,'' each of
which is  a maximal characterization of a system S in terms of the
experiments that can be performed upon it and  their possible
outcomes, subject to the constraints of internal consistency and
obedience to general physical laws and specific constitution. In the
case of the extremely simple system S each {\it candidate possible
world\/} is one of the following sixteen, each consisting of four
bits of information:   (L1$+$,  R1$+$).  (L1$+$, R1$-$), (L1$+$,
R2$+$), (L1$+$, R2$-$), (L1$-$, R1$+$), (L1$-$, R1$-$) (L1$-$,
R2$+$), (L1$-$, R2$-$), (L2$+$, R1$+$), (L2$+$, R1$-$), (L2$+$,
R2$+$), (L2$+$, R2$-$), (L2$-$, R1$+$), (L2$-$, R1$-$), (L2$-$,
R2$+$), (L2$-$, R2$-$). The qualification ``candidate'' is used
because predictions (1), (2), and (3) have not yet been applied to
winnow out impossibilities from the foregoing list. The list of
candidates for possible worlds is similar to the set of state
descriptions in the works of Carnap~\cite{21}, except that the
quantum mechanical restrictions against simultaneous performance of
experiments R1 and R2 and simultaneous performance of experiments L1
and L2 have limited the catalogue. Predictions (1)-(3) of the HER
model enforce the elimination of  (L1$-$, R2$-$),  (L2$+$, R1$+$),
(L2$-$, R2$+$) from the sixteen {\it candidate possible worlds},
leaving only thirteen {\it possible worlds\/}: (L1$+$,  R1$+$).
(L1$+$, R1$-$), (L1$+$, R2$+$), (L1$+$, R2$-$), (L1$-$, R1$+$),
(L1$-$, R1$-$) (L1$-$, R2$+$), (L2$+$, R1$-$), (L2$+$, R2$+$),
(L2$+$, R2$-$), (L2$-$, R1$+$), (L2$-$, R1$-$), (L2$-$, R2$-$). There
is quantitative information about probabilities in prediction (4),
information which does not restrict the set of possible states for
each system in the ensemble but only influences the statistics of
their actualization in the ensemble (each member of which is assigned
its own pair of space-like separated regions like the {\bf R} and
{\bf L} assigned to S). The analysis of counterfactual conditionals
for a single system does not depend upon the quantitative
probabilities but only upon the more primitive matter of possibility
or impossibility. Specifically, prediction (4) implies that (L1$-$,
R1$+$) and (L1$-$, R1$-$) are both possible worlds.

Finally I shall state two Auxiliary Assumptions which Stapp never
makes explicit but seems to be implicit in the HER quantum state and
in his proof of Property (II) (Stapp~\cite{1}, second column of p.\
31): (i) that the probability of R2$+$ is  greater than zero, given
that R2 is chosen to be performed in {\bf R} and that either L1 or L2
is chosen to be performed in {\bf L}; and (ii) that among the
ensemble of systems considered there are some in which L2 and R2 are
the chosen experiments, with outcome $+$ for R2, and some in which L1
and R2 are the chosen experiments,  with outcome $-$ for R2. These
assumptions guarantee that in the actual world for the ensemble some
systems have features which allow  Stapp's Property (II) to be proved
rigorously in Sect.\ 3, though---as will be seen---the  conjunction
of Properties (I) and (II) does not suffice to complete the proof of
the theorem at which he is aiming. The assumptions are of a kind
which is commonplace in the construction of interesting quantum
mechanical models. It should be emphasized that Assumption (ii) is
not a tautology unless quantum mechanical probability is interpreted
in a frequency sense, whereas a propensity interpretation is
convincingly proposed by Popper~\cite{22}) In practice, however,
Assumption (ii) is indispensable for the purpose of connecting
probability in the propensity sense  to observable frequencies.

\section{Stapp's Theorem}

The first stage in Stapp's demonstration of the theorem stated in
italics in the last paragraph of  Sect.\ 1 is a proof in column 2 of
p.\ 32 of the quantum theoretical Property (I) of the (HER) state:
that {\it if L2 is performed in an experiment of this type then  SR
necessarily follows, where SR asserts, ``If R2 is performed and gives
outcome $+$, then if, instead, R1 had been performed the outcome
would have been $-$.''}

There are two troublesome expressions in the definition of SR, namely
``if, instead'' and ``would have been.'' Stapp proposes a
clarification of the first of these expressions in the course of
proving Property (I):  ``The concept `instead' is given an
unambiguous meaning by the combination of the premisses of `free
choice,' and `no backward in time influence [NBITI]:'  the choice
between R1 and R2 is to be treated, within the theory, as a free
variable, and switching between R1 and R2 is required to leave any
outcome in the earlier region {\bf L} undisturbed. But then
statements (1) and (2) can be joined in tandem to give the result
SR.''   (Stapp~\cite{1}, column 2 of p.\ 31.)

The last sentence of this quotation is cryptic, but it is illuminated
by the discussion of ``would have been'' in column 1 of  p.\ 32 of
Stapp~\cite{1}. ``The previous argument rests heavily on the use of
counterfactuals:  the key statement SR involves, in a situation in
which R2 is performed and gives outcome $+$, the idea 'if, instead,
R1 had been performed\ldots\'' But then he says that Bell-type hidden
variables assumptions are also counterfactual, and he takes pains to
repeat that his assumptions are weaker than those of Bell and his
followers. Later on this page, in Section IV, he goes on to say that
his earlier exposition in Stapp~\cite{6} has been criticized, and
that even though he has answered these criticisms those challenges
indicate there are serious problems in that exposition, ``which
originate in the fact that it is based on classical modal logic.''
That logic, he says, has three drawbacks:
\begin{enumerate}
\item
Although the symbolic proof is concise and austere, that brevity is
based on a background that most physicists lack, which means that
most physicists cannot fully understand it without a significant
investment of time.
\item
The question arises as to whether the use of classical modal logic
begs the question by perhaps being based in implicit ways on the
deterministic notions of classical physics.
\item
Classical modal logic itself is somewhat of an open question, and it
is not immediately clear to what extent these issues undermine the
proof.
\end{enumerate}

Drawback (1) should be disregarded for two reasons. One is the
historical fact that physicists have learned to use much more
intricate mathematics than modal logic when that turned out to be
useful for physical problems. The other is that the formal proof on
p.\ 302 of Stapp~\cite{6} is in fact not concise and austere, since
its fourteen steps are susceptible to condensation---as shown in the
first paragraph on p.\ 850 of Shimony and Stein~\cite{14}  and the
second column of p.\ 31 of Stapp~\cite{1}.

Drawback (2) is troublesome on first inspection, since classical
modal logic does rely upon causal analysis, and the shift from the
deterministic laws of classical physics to the indeterministic laws
of QM would presumably require a modification of causal analysis.
However, an examination of Stapp's proofs of Property (I) and
Property (II) shows that only special cases of quantum mechanical
probabilities  are used:  impossibility and necessity in property
(I), that is to say, probability zero and unity, and necessity and
intermediate probability (neither zero nor unity) in property (II).
The ubiquitous statistical character of quantum mechanical
predictions, permitting the entire range of probabilities, has not
been invoked.

Drawback (3) is not troublesome because of an important contribution
by Stapp himself. In the often-cited work on the logic of
counterfactuals by D. Lewis~\cite{23} there is indeed an open
question: how to give a reasonable criterion of the comparative
closeness of possible worlds w$'$ and w$''$ to actual world w, a
criterion that is needed to implement Lewis's truth condition for a
counterfactual conditional proposition. The discussion in Lewis's
book sounds quite scholastic, and one has a feeling that his entire
program is endangered by the likelihood that the question of
comparative closeness is not well posed. Stapp cuts through the
question by proposing that {\it a counterfactual conditional ``if p
were true, then q would be true'' is true if and only if q is true in
every possible world w$'$ that differs from the actual world w only
by the consequences of the action described by p; according to STR
this condition means every possible world w$'$ in which p is true and
which agrees with w everywhere outside the future light-cone of the
set of space-time points where the experimenter's action is
localized\/} (paraphrase of paragraph 5 in column 1 on p.\ 855 of
Stapp~\cite{8}); also Shimony and Stein~\cite{15}, paragraph 2 on p.\
502). A clear criterion from STR has thus replaced the elusive
criterion of ``closeness'' of one world to another. Stapp has thus
admirably solved the problem that has made him wary, in drawback (3),
of using classical modal logic.

With these clarifications of counterfactual conditional propositions
we can prove Stapp's Property (I), which asserts (in column 2 of p.\
31 of Stapp~\cite{1}), ``if an experiment of the Hardy-type is
performed then L2 implies SR, where SR $=$ `If R2 is performed and
gives outcome $+$, then if instead R1 had been performed the outcome
would have been $-$'.'' This conclusion is correct even if
``implies'' is taken in its strongest sense,  the sense of {\it
strict implication (i.e., ``p strictly implies q'' is true  if and
only if in every possible world in which p is true q is also true)}.
The possible worlds, as stated in Section 2, are {\it maximal
characterizations of the system S consistent with each each other and
with the obedience by S of the general laws of physics (the relevant
ones being the laws of quantum mechanics) and the specific
constitution of S (the HER state): The antecedent of the implication
asserted by property (I) is that L2 is performed in {\bf L}, which is
a premiss that picks out six of the thirteen possible worlds\/}:
(L2$+$, R1$-$), (L2$+$, R2$+$), (L2$+$, R2$-$), (L2$-$, R1$+$),
(L2$-$, R1$-$), (L2$-$, R2$-$). The antecedent of SR is false in all
of these except (L2$+$, R2$+$), and therefore SR itself is trivially
true in these five cases by the logic of material implication. But in
the case of the world (L2$+$, R2$+$) the consequent of SR is true,
since the only possible world in which R1 is performed instead of R2
and which is in agreement with the part of (L2$+$, R2$+$) in region
{\bf L} is (L2$+$, R1$-$). Hence SR is true in all possible worlds in
which L2 is performed, ensuring the truth of Property (I).

The second stage in Stapp's demonstration of the theorem stated in
Sect.\ 1 is to prove Property (II), which asserts, ``Quantum theory
predicts that if an experiment of the Hardy-type is performed then
`L1 implies SR' is false.'' This assertion is correct if ``implies''
is interpreted in the sense of strict implication. The premisses of
Property (II) are the correctness of quantum mechanics, the HER
characterization of S, and the performance of L1 on $\lambda$. By
Assumption (ii) there are members of the ensemble in which L1 is
actually performed. One must check that SR holds of the part in {\bf
R} of any such member. If R2 is either not performed on the part of
the system in {\bf R} or is performed with outcome---then SR is
trivially true by the logic of material implication, because its
antecedent is false. On the other hand, if in the actual world R2 is
performed on this system with outcome $+$,  then SR is true of the
system if and only if the consequent of SR is true, i.e., if and only
if R1 has outcome---in all possible worlds in which R1 is performed
on the part of this system in {\bf R} but which agree with the actual
world outside the forward light cone of  {\bf R}, in particular in
region {\bf L}. By Assumption (ii) there are systems in the ensemble
for which L1 is actually performed on the members in {\bf L} with
outcome---while experiment R2 is freely chosen to be performed on the
corresponding systems in {\bf R} By prediction (3) R2 has outcome $+$
for these systems, thus guaranteeing the truth of the antecedent of
SR. By prediction (4) in each  possible world in which R1 is
performed instead of R2 for the systems just mentioned there is a
non-zero probability that the outcome will  be R1$+$. Hence (L1$-$,
R1$+$) is a possible world. Consequently there is a case in which the
consequent of SR is false even though the antecedent of SR is true.
Hence we have proved that {\it there is a  case in which L1 is the
experiment chosen in {\bf L} but SR is false, thus establishing
Property (II)}.

Stapp then combines Properties (I) and (II) to conclude that the free
choice between L1 and L2 has a causal relation to the truth of the
proposition SR. He claims that ``in any theory or model in which the
three assumptions (free choice, NBITI, certain predictions of quantum
theory ) are valid, the statement SR must always be true if the free
choice in region {\bf L} is L2, but must sometimes be false if that
free choice in {\bf L} is L1. But the truth or falsity of SR is
defined by conditions on the truth or falsity of statements
describing possible events located in region {\bf R}. The fact that
the truth of S depends in this way on a free choice made in region
{\bf L}, which is space-like-separated from region {\bf R}, can
reasonably be said to represent the existence within the theory or
model of {\it some sort\/} of faster-than-light influence.''
(Stapp~\cite{1}, bottom of p.\ 31, top of p.\ 32.)

The error in Stapp's argument is his claim that SR is a statement
about region {\bf R} alone. To be sure, the only {\it events\/}
mentioned explicitly in SR are choices between experiments in {\bf R}
and outcomes of the chosen experiments. But SR is not simply a
statement about actually occurring events. It is a counterfactual
conditional, and its truth  condition is (by substituting in the
italicized passage in Sect.\ 2) the following: {\it R1$-$ is true in
every possible world w$'$ that differs from the actual world w only
by the consequences of the action described by R1;  according to STR
this means every possible world w$'$ in which R1 is true  and which
agrees with w everywhere outside the forward light-cone of {\bf R}}.
The phrase ``outside the forward light-cone of {\bf R}'' applies to
{\bf L} because of the assumption that {\bf R} and {\bf L} are
space-like separated. Hence SR does refer to parts of spacetime
outside the region {\bf R}. And once  this characteristic of SR is
recognized one sees immediately that the choice between L1 and L2 in
region {\bf L} is not a superluminal cause of an effect in the
space-like separated region {\bf R}, causing SR to be true if the
latter choice is made and SR to be false if the former choice is
made. SR is an intricate proposition involving three different kinds
of entities: events in {\bf R}, namely the actual choice between
experiments R1 and R2 for $\rho$ and the outcome of the chosen
experiment;  events in {\bf L}, namely the actual choice between
performing experiment L1 or L2  on $\lambda$ and the outcome of the
chosen experiment;  and a set of possible worlds satisfying the
requirements spelled out in the truth condition for the
counterfactual conditional that constitutes the consequent of SR,
this set being  an entity determined jointly by the aforementioned
events in {\bf L} and the aforementioned events in {\bf R}. Stapp's
assertion that SR is localized in region {\bf R} is clearly
incompatible with the intricacy of SR.

An answer to the foregoing argument is given in Stapp~\cite{8}, p.\
857, column 2, paragraph 2: ``My input-output analysis makes the
following point:  The fact that (1), the truth or falsity of this
statement SR is, for any fixed choice made by the experimenter in
{\bf L}, determined explicitly by whether or not a certain
conceivable event, R1$-$, must occur in {\bf R} under conditions
defined in {\bf R}, coupled with the agreed-upon fact that (2) the
truth of SR depends upon which choice is made by the experimenter in
{\bf L}, means that whether or not this conceivable event in {\bf R}
must occur depends upon which choice is made by the experimenter in
{\bf L}.'' Stapp misstates the conditions under which SR is true or
false. They are indeed determined by whether or not a certain
conceivable event, R1$-$, must occur in {\bf R} under certain
conditions, but it is not the case that these are ``conditions
defined in {\bf R}.'' Yes, two of the conditions are indeed defined
in {\bf R}: the actual performance of R2 with outcome $+$, and the
restriction of relevant possible worlds in which the outcome R1$-$ is
required to those in which R1 is performed instead of R2;  {\bf but
the further restriction upon the relevant possible worlds, that they
agree with the actual world outside the forward light-cone of {\bf
R}, is not a condition defined in {\bf R}, because the region in
which agreement is enforced includes {\bf L}}.

An analogy may provide some relief from the logical complexity of the
preceding paragraph. Some properties can be attributed to a local
region {\bf R} of space-time intrinsically, without any reference to
other space-time regions. Others, however, cannot be attributed
intrinsically. For instance, a comparative or superlative property p
like ``locus of the fastest hundred meter free-style swim'' cannot be
attributed to {\bf R} without making a comparison with other regions.
Indeed, if the choice is freely made to schedule a hundred meter
free-style swimming match in a region {\bf R$'$} in the forward
light-cone of {\bf R} and the outcome is a world's record, then p
will not hold of {\bf R}, but it  cannot be (note NBITI) that an
event in {\bf R$'$} is the cause of an event in {\bf R}.  This
remark, of course, is a banality. But some sophistication and
reflection is required in order to understand that a modal property
like SR has a relational character just as much as the manifestly
superlative property p.

\section{Conclusion}

We can now see the virtue of the demonstrations of Bell's theorem
offered by Bell~\cite{16,17} himself and by his followers~\cite{18},
which assume a hidden-variables model as a premiss.  Such a
model---as Einstein, Podolsky, and Rosen~\cite{24}
understood---explains correlations between regions {\bf R} and {\bf
L} by means of attributions of intrinsic properties (their ``elements
of physical reality'') to each region. Consequently the correlations
cannot be dismissed as banalities stemming from the relational
character of the attributed properties. The premisses of
demonstrations offered by Bell and his followers are indeed stronger
than those offered by Stapp, but the reward of the stronger premisses
is a logically impeccable conclusion that certain experimentally
demonstrated quantum mechanical correlations violate relativistic
locality.

In column 2 of p.\ 32 of Stapp~\cite{1} one finds an interesting
historical remark, ``The EPR argument rests strongly on
counterfactual ideas.'' This statement is widely held, and I myself
believed for a long time (Shimony~\cite{25}) that  a commitment to
counterfactual reasoning is a corollary of EPR's physical realism.
Recently, however, I have been convinced by an old thesis of
d'Espagnat~\cite{26} that EPR could reach their conclusions by
ordinary inductive logic, without any invocation of counterfactual
conditionals.  Shimony~\cite{27}  presents an elaboration of
d'Espagnat's position  and argues, in addition, that EPR's ``elements
of physical reality'' would suffice to provide a grounding for
counterfactual conditionals. Of course, the preceding sentence is
itself conditional, and it is undermined by Bell's theorem and the
related experiments. These throw doubt on the assumption of
relativistic locality needed by EPR to reach their conclusion that
there exist independent but correlated elements of physical reality
in spacelike separated regions.

\section*{Acknowledgement}

The author is indebted to Howard Stein for past collaborations in
which he clarified the logic of counterfactual conditionals.

\end{document}